\documentclass[12pt]{article}
\usepackage{latexsym}

\def\cH{{\cal H}}
\def\cR{{\cal R}}
\def\cG{{\cal G}}
\def\cS{{\cal S}}
\def\cD{{\cal D}}

\def\tr{{\rm tr}}
\def\ket#1{\mid~\!\!\!{#1}~\!\!\rangle}
\def\bra#1{\langle~\!\!{#1}~\!\!\!\mid}

\def\sign{\buildrel + \over -}
\def\N{_{1\dots N}}

\def\6{\Big[(N!)\Big/\prod_{j=1}^J(N_j!)
\Big]}
\def\S{S_{1\dots N}^{s,a}}
\def\M{_{(M_j+1)\dots (M_j+N_j)}}

\def\QM{{\rm quantum mechanics }}
\def\qm{{\rm quantum mechanics}}
\def\${\enspace$}

\begin{document}

\begin{center}
{\bf \Large \noindent HOW TO
DISTINGUISH IDENTICAL PARTICLES.
THE GENERAL CASE}\\

\large \rm FEDOR HERBUT

{\it \footnotesize  Serbian Academy of
Sciences and
Arts, Knez Mihajlova 35,\\
11000 Belgrade, Serbia\\
fedorh@infosky.net and
fedorh@mi.sanu.ac.yu}\\

\end{center}

\footnotesize \noindent The
many-identical-particle quantum
correlations are revisited utilizing
the machinery of basic group theory,
especially that of the group of
permutations. It is done with the
purpose to obtain precise definitions
of effective distinct particles, and of
the limitations involved. Namely,
certain restrictions allow one to
distinguish identical particles in the
general case of \$N\$ of them, and of
\$J\$ clusters of effectively distinct
identical particles, where \$N\$ and
\$J\$ are arbitrary integers (but
\$2\leq J\leq N\$). Mutually
orthogonal, single-particle
distinguishing projectors (events or
properties), \$J\$ of them, are the
backbone of the construction. The
general results are exemplified by
local \qm, and by the case of nucleons.
The former example suits laboratory
experiments, and a critical view of it
is presented.\\


\normalsize\rm

\section{Introduction}

\noindent The inventor of the exclusion
principle, Pauli, is reported to have
said (private communication by the late
R. E. Peierls) that if two electrons
are apart, then they are distinct
particles by this very fact. His
principle applies to those that are not
in this relation. De Muynck has pointed
to$^1$ Schiff's unsuccessful
formalization$^2$ of Pauli's statement.

Generalizing Pauli, Schiff
stipulates$^2$ that two {\it identical
particles are distinguishable} when the
two-particle probability amplitude
\$a(1,2)\$ of some dynamical variable
is different from zero only when the
two particles have their values in {\it
disjoint} ranges of the spectrum of the
variable. But, as de Muynck
remarks,$^1$ this actually {\it cannot
ever occur} when the wave function is
(anti)symmetric, for then
\$a(2,1)=\sign a(1,2)\$ (for identical
bosons and identical fermions
respectively).

In the next section it will be shown
that the present author's previous
result$^3$ for two identical particles
in the form of a suitable theorem
resolves this difficulty.

It will also be explained how this
theorem incorporates Mirman's important
claim$^4$ that distinguishability of
identical particles is essentially an
experimental notion. Namely, following
Jauch,$^5$ one can distinguish {\it
intrinsic and extrinsic properties} of
particles. According to him, identical
are those particles that have equal
intrinsic properties. But, as de Muynck
remarks,$^1$ "an intrinsic property may
show up dynamical behavior", and turn
out to be extrinsic like the proton and
neutron states (cf subsections 5.C and
6.C). It all depends on the
experimental conditions.

In Section 3 some mathematical notions
required for the intended
generalization to \$N\$ particles are
presented. In Section 4 the mentioned
theorem from previous work,$^3$ which
makes it clear how one can distinguish
two identical particles turning
formally extrinsic properties into
intrinsic ones, is generalized to an
arbitrary number of particles and an
arbitrary number of effectively
distinct clusters of particles.

The analysis is set against quotations
from two standard textbooks on quantum
mechanics, that of Messiah,$^6$ and
that of Cohen-Tannoudji et al.$^7$ An
attempt is made to show where these
textbooks are right and where they lack
detail and precision in an important
way.

In Section 5 some illustrations are
given. Finally, in Section 6 concluding
remarks point to the salient features
of the article.\\

In {\it first-quantization} \QM one has
\$N\$ single-particle state spaces
\$\{\cH_n:n=1,\dots ,N\}.\$The
identicalness of the particles is
expressed in terms of isomorphisms
\$\{I_{i\rightarrow j}:i,j=1,\dots
,N;i\not= j\}\$ connecting pairs of
single-particle spaces: \$
\cH_j=I_{i\rightarrow j}\cH_i,\enskip
i,j=1, \dots ,N;\enskip i\not= j.\$
Naturally, \$I_{i\rightarrow
j}I_{j\rightarrow i}=I_j,\$ \$I_j\$
being the identity operator in
\$\cH_j.\$

As an illustration for the action of
the operators \$I_{i\rightarrow j}\$ we
mention that the second-particle
radius-vector operator is: \$\vec
r_2=I_{1\rightarrow 2} \vec r_1
I_{2\rightarrow 1}.$

The {\it effective N-distinct-particle
space}, on which the description of
identical particles in
first-quantization \QM is based, is \$
\cH_{1\dots N}\equiv
\prod_{n=1}^{\otimes N}\cH_n, \$ where
\$\otimes\$ denotes the tensor (or
direct) product of Hilbert spaces. (We
shall use this symbol also for the
tensor product of vectors and of
operators.)

The basic mathematical tool in the
investigation that follows is
elementary group theory, in particular,
the use of the group \$\{p:p\in
\cS_N\}\$ of all permutations \$p\$ of
\$N\$ objects. (This group is usually
called "the symmetric group". We will
use the usual symbol \$\cS_N,\$ but not
the term to avoid confusion because it
is not the group that is symmetric, but
the group is the natural tool to
express the identical-particle
symmetries of the state of the system.)

To make the reading easier for those
theoretical physicists who are not
quite familiar with group theory, not
even in its elementary form, the
requisite group-theoretic machinery is
systematized, and to some extent
derived, in Appendix A. Proof of
Theorems 1 and 2, which are hopefully
new and are the main result (but their
proof is somewhat more intricate) is
relegated to Appendices B and C
respectively. Finally, in Appendix D
the concept of possession of a property
by a system in a state and by an
observable is explained.\\

\section{Distinguishability of two
identical particles}

\noindent As it was mentioned, the
identical-particle idea rests on
distinguishing intrinsic and extrinsic
properties of a particle. The former do
not enter the quantum-mechanical
formalism; they form the physical basis
of the single-particle state space (e.
g., of that of the electron, which is
characterized by mass, charge, spin,
gyromagnetic factor etc, as unique
intrinsic properties). The extrinsic
properties play an important role in
the formalism in terms of projectors.

Returning to Schiff's attempt to
formalize a generalization of Pauli's
distinguishing identical particles$^2$
(cf the Introduction), we imagine that
the single-particle observable at
issue, if incomplete, is completed by
some suitable compatible observables
into a complete set (in principle, this
is always possible), and that the
two-particle amplitude \$a(1,2)\$ is
the two-particle wave function in the
representation of this complete set.
Then, we learn from textbooks that
\$a(2,1)=\sign a(1,2)\$ for identical
bosons and identical fermions
respectively.

Let the projectors \$E\$ and \$F\$
correspond to two {\it disjoint
regions} in the spectrum of the
observable at issue. Introducing the
orthocomplementary projector
\$E^{\perp}\enskip\Big(\equiv
1-E\Big)\$ (which, in this case,
projects onto the set-theoretical
complementary region in the spectrum),
the relation \$E_1\phi(1)=\phi(1),\$ is
equivalent to \$E_1^{\perp}\phi(1)=0.\$
Hence, the former relation (as also the
latter) means that for particle \$1\$
the observable at issue has positive
probability values only in the
corresponding region. Analogous
statements hold true for \$F.\$

Let, further, the index value in
\$E_i,\$ \$F_i,\$ \$i=1,2\$ show to
which of the particles the extrinsic
property applies. Then, {\it the
correct way to express Pauli's
criterion of distinguishability} is to
say that the two-particle system
possesses the property
\$(E_1F_2+F_1E_2)\$ (it must be
symmetrized): \$(E_1F_2+F_1E_2)a(1,2)=
a(1,2).\$(More on the concept of
possession of a property in Appendix
D.)

The mathematical results of previous
work,$^3$ in terms of isomorphism and
equivalence of relevant operators, then
show that the extrinsic properties
\$E\$ and \$F\$ can be transformed
effectively into intrinsic ones by {\it
isomorphic transition} from the
subspace \$(E_1F_2+F_1E_2)S_{12}^{s,a}
(\cH_1\otimes\cH_2)\$ (\$S_{12}^{s,a}\$
denotes the symmetrizer or the
antisymmetrizer) to the effective
distinct-particle state space
\$\Big(E_1\cH_1 \otimes
F_2\cH_2\Big).\$ Schiff's mentioned
criterion is actually  valid in the
latter, distinct-particle space.

Mirman's claim$^4$ of the essential
role played by experiments shows up in
the fact that the mentioned
transformation of extrinsic properties
into effective intrinsic ones is
restricted to experiments in which the
possession of the property
\$(E_1F_2+F_1E_2)\$ is preserved.

Thus, a generalized Pauli criterion of
distinguishing identical particles can
be expressed in the quantum-mechanical
formalism quite satisfactorily as far
as two identical particles are
concerned. The motivation for this
article is the belief that
generalization to any number of
particles and any number of
distinguishing properties is desirable.
The more so because there are two
important examples of effective
distinguishing identical particles:
that of non-overlapping spatial domains
(see subsection 5.B), and the case of
nucleons in nuclei (subsections 5.C and
6.C). \\

\section{Identical particles and the
maximal-symmetry projectors}

{\bf\noindent Definition 1.} One speaks
of identical particles if the particles
have identical complete sets of
intrinsic properties.

This condition has the prerequisite
that long experience suggests that one
is unable to convert any of the
intrinsic properties by dynamical means
into extrinsic ones, and that one is
unable to extend the set of such
properties. These are {\it impotency
stipulations} analogous to those of
thermodynamics on which the
thermodynamical principles are based.\\

Explanation is in order. Some time ago
the electron neutrino and the muon
neutrino were believed to be identical
particles because they had their,
up-to-then known, intrinsic properties
in common. Later it was discovered that
they differ; the former has the
electronic leptonic quantum number, and
the latter the muonic one. Thus, their
other common properties were
incomplete; after completion it turned
out that they no longer have all
intrinsic properties equal.

An illustration for converting an
intrinsic property into an extrinsic
one is the case of parity and weak
interaction. Until the advent of the
famous parity-non-conserving weak
interaction experiments, parity could
be considered an intrinsic property of
the elementary particles. These
experiments converted it into an
extrinsic one, and nowadays we must
work with the parity observable with
its parity-plus and parity-minus
eigen-projectors.\\

Now, a few remarks of mathematical
nature on how one associates the
unitary operator representative \$P\N\$
in \$\cH\N\$ with a given permutation
\$p\in \cS_N.\$

If \$p\$ is a transposition,
transposing, e. g., the state vectors
of the first and the second particle,
then the corresponding permutation, a
so-called {\it exchange operator}, acts
on uncorrelated \$N$-particles vectors
as the operator tensor product \$
P_{1\dots N}^{1\leftrightarrow 2}
\equiv I_{1\rightarrow 2}\otimes
I_{2\rightarrow 1}\otimes I_3 \otimes
\dots \otimes I_N\$ (cf A.III.1 in
Appendix A). Its action on correlated
vectors follows then uniquely as an
immediate consequence of requiring
linearity and continuity. It is easily
seen that this operator takes an
uncorrelated basis (induced from the
factor spaces) into itself because it
amounts to a permutation of its
vectors. Hence the operator is unitary.

All transpositions (special
permutations) in \$\cH_{1\dots N}\$ are
analogously connected with the
corresponding isomorphisms
\$I_{i\rightarrow j},\$ and, as it is
well known, all permutations factorize
into transpositions. In this manner,
all permutations in \$\cH\N\$ can be
defined in terms of the isomorphisms \$
\{I_{i\rightarrow j}:i,j=1,\dots
,N;i\not= j\} \$ connecting pairs of
single-particle state spaces. All
permutation operators are unitary
because any product of unitary
operators is unitary.

It is well known that bosons and
fermions differ sharply in some
properties (e. g., Bose condensation).
But, as it was shown in previous
work,$^3$ as far as {\it
distinguishing} identical ones of them
goes, they behave equally. In this
study we extend this result to any
number of particles.

To treat bosons and fermions together,
we will write \$ sign(p),\$ which is,
by definition, \$1\$ if one treats
bosons, and it equals \$(-)^p,\$ the
parity of the permutation (cf A.II.1)
if one deals with fermions.

Now we write {\it the symmetry theorem}
in a concise and practical form. For
every permutation \$ p\in \cS_N\$ and
every state vector \$\ket{\psi }\N\$ of
an \$N$-identical-particle system the
action of the former on the latter
amounts to no more than a possible
change of sign as follows:
$$P\N\ket{\psi }\N =
sign(p)\ket{\psi }\N\eqno{(1a)}.$$

It is known that the identical-particle
symmetry correlations are expressible
in terms of {\it maximal-symmetry
operators}. We define them as
projectors. They are \$S^{s,a}_{1\dots
N} \equiv \Big(S^s_{1\dots N} \enskip
\mbox{or}\enskip S^a_{1\dots N}\Big).\$
The {\it symmetrizer} (for identical
bosons) is \$ S^s_{1\dots N} \equiv
(N!)^{-1} \sum_{p\in \cS_N}P_{1\dots
N},\$ and the {\it antisymmetrizer}
(for identical fermions) is \$
S^a_{1\dots N}\equiv (N!)^{-1}
\sum_{p\in \cS_N}(-1)^pP_{1\dots N},.\$
Thus, \$ S^{s,a}_{1\dots N}\$ is the
\$N$-identical-particle {\it
maximal-symmetry projector operator}.
We write
$$S^{s,a}_{1\dots N}=(N!)^{-1}\sum_{p\in
\cS_N}sign(p)P_{1\dots N}. \eqno{(2)}$$

Maximal symmetry (boson symmetry or
fermion antisymmetry) of a state vector
can be expressed, besides by (1a), also
(equivalently) by $$S\N ^{s,a}\Psi \N
=\Psi \N\eqno{(1b)}.$$ It is
straightforward to prove this claim.

The geometrical meaning of (1b) is that
every physically meaningful state
vector is within the subspace
\$S^{s,a}\N\cH\N\$ of \$\cH\N.\$We call
the former the {\it first-principle
state space} of identical particles.

Obviously, each mixed-or-pure state
(density operator) \$\rho\N\$ has its
range within the former subspace, or,
equivalently, any decomposition of
\$\rho\N\$ into pure states results in
state vectors from the subspace
\$S^{s,a}\N\cH\N.\$In standard
language, one speaks of Bose-Einstein
statistics if one has bosons (if
\$S^{s,a}\N = S^s\N\$), and of
Fermi-Dirac statistics in the case of
fermions (when \$S^{s,a}\N =S^a\N\$).\\

\section{How to obtain distinct particles}

\noindent In both textbooks
Cohen-Tannoudji et al.$^7$ and
Messiah$^6$ the way how to distinguish
identical particles is presented in
some detail and fairly correctly (cf
pp. 1406-1408 in the former and pp.
600-603 in the latter). For instance,
in Messiah (pp. 600-601) one can find
the following passage.

\leftskip 9ex \noindent \footnotesize
"In practice, the electrons of a system
are all inside a certain spatial domain
\$\cD,\$ and the dynamical properties
in which we are interested all
correspond to measurements to be made
inside this domain. It turns out that
the other electrons may simply be
ignored so long as they remain outside
\$\cD\$ and so long as their
interaction with the electrons of the
system remain negligible. This is a
general result and applies to bosons as
well as to fermions. We shall prove it
here for the special case of a system
of two fermions."

\leftskip 0cm \normalsize The exposition is restricted to spatial
and spin-projector distinctions. But the general procedure of
distinguishing is not given, and the precise restrictions involved
are not clear. It is the purpose of the two theorems that
follow to make up for these deficiencies.

We utilize the powerful tool of
projectors. Thus, \$ Q_{\cD}\equiv \int
\int \int_{\cD}\ket{\vec r} \bra{\vec
r} d^3\vec r,\$ and \$ Q_{out}\equiv
\int \int \int_{\cD^c} \ket{\vec
r}\bra{\vec r} d^3\vec r,\$ are the
projectors corresponding to the
mentioned domain \$\cD\$ and to the
complementary (in the set-theoretical
sense) domain \$\cD^c\$ in {\bf
R}$_3,\$ which means "outside \$\cD\$".
One should note that \$Q_{\cD}\$ and
\$Q_{out}\$ are orthogonal:
\$Q_{\cD}Q_{out}=0,\$ and that
\$Q_{out}=Q_{\cD}^{\perp}\$
(\$Q_{\cD}^{\perp}\$ being the
projector orthocomplementary to
\$Q_{\cD}\$). Thus, in the quoted
passage, these two projectors
distinguish between the electrons that
one is interested in and
those that one is not.\\

{\it In the general case}, which we are
now going to investigate, let the {\it
distinguishing properties or events} be
given by arbitrary \$J\$ orthogonal
single-particle projectors:
\$\{\{Q_n^j:j=1, \dots ,J\}:n=1,\dots
,N\},\$ \$\forall j, \forall n:\enskip
(Q_n^j)^{\dagger}=Q_n^j\$ (Hermitian
operators), \$\forall n:\enskip
Q_n^jQ_n^{j'}= \delta_{j,j'}Q_n^j\$
(orthogonal projectors), and finally,
\$\forall j:\enskip Q^j_n=
I_{1\rightarrow n}Q_1^jI_{n\rightarrow
1}, \enskip n=2,\dots ,N\$
(mathematically, equivalent projectors;
physically, same properties or events).

We have in mind \$J\$ clusters of
effectively-distinct particles ,
\$2\leq J\leq N.\$We write them in an
{\it ordered} way according to the
(arbitrarily fixed) values of \$j:\$
\$j=1,\dots ,J.\$The \$j$-th cluster
contains a certain number of particles,
which we denote by \$N_j,\$ \$
\sum_{j=1}^JN_j=N.\$It will prove
useful to introduce also the sum of
particles up to the beginning of the
\$j$-th cluster: \$ M_j\equiv
 \sum_{j'=1}^{(j-1)}N_{j'}\$ for \$j\geq 2,\$
 and \$M_1\equiv 0.\$

The distinguishing projectors appear in
\$\cH\N\$ through the {\it tensor
product of distinguishing projectors}:
$$Q_{1\dots N}\equiv
\prod_{j=1}^{\otimes J}
\Big(\prod_{n=(M_j+1)} ^{\otimes
(M_j+N_j)}Q_n^j\Big).\eqno{(3a)}$$ One
should note that the last product (in
the brackets) applies to the \$j$-th
{\it cluster}, and that it multiplies
tensorically physically equal
(mathematically equivalent via
transpositions) single-particle
projectors.

We introduce the corresponding
effective {\it distinct-cluster space}
\$\cH_{1\dots N}^D,\$ which is the
state space of \$J\$ ordered
distinct-particle clusters, each
consisting of identical particles:

$$\cH_{1\dots N}^D\equiv \Big\{
\prod_{j=1}^{\otimes J} \Big[S\M
^{s,a}\Big(\prod_{n=(M_j+1)} ^{\otimes
(M_j+N_j)}Q^j_n\Big)\Big]\Big\}\cH\N=$$
$$\prod_{j=1}^{\otimes J}
\Big[S\M ^{s,a}\Big(\prod_{n=(M_j+1)}
^{\otimes
(M_j+N_j)}(Q^j_n\cH_n)\Big)\Big]=$$
$$
Q_{1\dots N}\Big(\prod_{j=1}^{\otimes
J}S_{(M_j+1)\dots (M_j+N_j)}^{s,a}
\Big)\cH_{1\dots N},\eqno{(4a,b,c)}$$
where \$a,b,c\$ refer to the three
expressions of \$\cH_{1\dots
N}^D.\$(and the two operator factors in
(4c) commute).

Note that the distinct {\it cluster
spaces} (factors in the tensor product
\$ \prod_{j=1} ^{\otimes J}\$ in (4a)
or (4b)) are {\it decoupled} from each
other (in the sense of
identical-particle symmetry
correlations), i. e., one has the
tensor product \$\prod_{j=1}^{\otimes
J},\$ but the factor spaces within each
cluster are coupled by the
corresponding maximal-symmetry
projectors.

On the other hand, there is the {\it
symmetrized tensor product of
distinguishing projectors} in \$\cH\N\$
determined by (3a) and the permutation
operators:

$$Q_{1\dots N}^{sym}\equiv \bigg(
\sum_{p\in \cS_N} \Big(P_{1\dots
N}Q_{1\dots N}P_{1\dots
N}^{-1}\Big)\bigg)\Big/
\prod_{j=1}^J(N_j!).\eqno{(3b)}$$ We
call it the {\it distinguishing
property}.

For each term in (3b) there exist \$
\Big([\prod_{j=1}^J(N_j!)]-1 \Big)\$
other terms equal to it (cf (3a)).
There are \$\Big(
(N!)\Big/[\prod_{j=1}^J(N_j!)]\Big)\$
distinct terms in (3b), and they are
orthogonal projectors in \$\cH\N.\$The
operator \$Q\N^{sym}\$ is a symmetric
projector, i. e., one that commutes
with every permutation operator
\$P\N.\$(Proof of these claims see in
A.III.4 and A.III.5.)

The {\it corresponding
\$N$-identical-particle subspace}
\$\cH_{1\dots N}^{Id}\$ of \$\cH\N\$ is
defined as the range of \$Q\N^{sym}\$
in the first-principle state space:
$$\cH_{1\dots N}^{Id}\equiv Q_{1\dots N}^{sym}
S^{s,a}_{1\dots N} \cH_{1\dots N}=
S^{s,a}_{1\dots N}Q_{1\dots N}^{sym} \cH_{1\dots N}
.\eqno{(5)}$$

{\bf\noindent Theorem 1.} {\it The
subspaces \$ \cH^{Id}_{1\dots N}\$ and
\$ \cH^D_{1\dots N}\$ are {\rm
isomorphic}, and the maps
$$I\N^{Id\rightarrow D}\equiv
\Big((N!)\Big/[\prod_{j=1}^J(N_j!)]
\Big)^{1/2}Q_{1\dots N}|_{\cH\N^{Id}},
\eqno{(6)}$$
$$I\N^{D\rightarrow Id}\equiv
\Big((N!)\Big/[\prod_{j=1}^J(N_j!)]
\Big)^{1/2} S^{s,a}_{1\dots
N}|_{\cH\N^D},\eqno{(7)}$$ where
\$|_{\dots }\$ denotes the restriction
to the corresponding subspace, are {\it
mutually inverse unitary isomorphisms}
mapping \$ \cH^{Id}_{1\dots N}\$ onto
\$ \cH^D_{1\dots N}\$ and vice versa:
\$ \cH\N^D=I\N^{Id\rightarrow
D}\cH\N^{Id}\$ and \$ \cH\N^{Id}=
I\N^{D\rightarrow Id}\cH\N^D.\$}

Theorem 1 is proved in Appendix B.\\

{\bf\noindent Definition 2.} In case
the state \$\rho\N^{Id}\$ of an
\$N$-identical-particle system
satisfies the relation
$$Q\N^{sym}\rho\N^{Id}
=\rho\N^{Id} ,\eqno{(8)}$$ we say that
the system {\it possesses} the {\it
distinguishing property}
\$Q^{sym}_{\N}\$ in the state in
question (cf relations (D.2) and (D.3)
in Appendix D). In this case, and only
in this case, it is amenable to Theorem
1. Since (8) is actually a {\it
restriction} on the choice of state, we
refer to \$Q\N^{sym}\$ also as
the {\it restricting property}.\\

The distinguishing (single-particle)
properties \$ \{Q^j:j=1, \dots ,J\}\$
determine the restricting property
\$Q\N^{sym}.\$ This is the backbone of
the presented answer to the question
"How to distinguish identical
particles?".

The physical meaning of the {\it
decoupling} and the {\it coupling
isomorphisms} \$I\N^{Id\rightarrow D}\$
and \$I\N^{D\rightarrow Id}\$
respectively given in the theorem shows
up, of course, in the {\it observables}
that are defined in \$\cH\N ^{Id}\$ and
\$\cH\N ^D.\$The corresponding or
equivalent operators (obtained by the
similarity transformation) are of the
same kind: Hermitian, unitary,
projectors etc. because all these
notions are defined in terms of the
Hilbert-space structure, which is
preserved by the (unitary)
isomorphisms. In Theorem 2 (on
observables) below, the restricting
role of \$Q\N^{sym}\$ will be
additionally clarified.\\

It is seen that a prerequisite for
describing an evolution or a
measurement in the subspaces \$\cH\N
^{Id}\$ and \$\cH\N ^D\$ is the
possession of the restricting
properties (occurrence of the events)
\$Q\N^{sym}\$ and \$Q\N\$ respectively,
and their preservation.

As it is clear from (5), a relevant
observable for the decoupling, i. e., a
Hermitian operator that reduces in
\$\cH\N ^{Id},\$ is one that {\it
commutes} with the restricting
projector \$Q\N ^{sym},\$ and {\it one
confines oneself to its reducee} in
\$\cH\N^{Id}\$ (cf (5)). In physical
terms, the observable must be {\it
compatible} with the restricting
property \$Q\N ^{sym}\$ and one must
assume that the property is {\it
possessed} (cf (D.2) and (D.3) in
Appendix D), and that this is preserved
if some process is at issue.\\

{\bf\noindent Theorem 2. A)} {\it Let
\$A\N^D\$ be any Hermitian operator
(observable) in \$\cH\N\$ that {\it
commutes} (is compatible) {\it both}
with every permutation associated with
the distinct-cluster representation
\$\forall p\in\cG_D:\enskip
[A\N^D,P\N]=0\$ (cf property A.III.4)
and with \$Q\N\$ (cf (3a)). (Hence,
\$A\N^D\$ reduces in \$\cH\N^D.\$) Let,
further,
$$A\N^{D,Q,sym}\equiv
\Big(\prod_{j=1}^J(N_j!)\Big)^{-1}
\sum_{p\in\cS_N}P\N A\N^DQ\N
P\N^{-1}\eqno{(9)}$$ be the symmetrized
product \$A\N^DQ\N.\$ Then (the
symmetric operator) \$A\N^{D,Q,sym}\$
commutes with \$Q\N^{sym}\$ (cf (3b)),
and the {\it reducee} of \$A\N^D\$ in
\$\cH\N^d\$ and that of
\$A\N^{D,Q,sym}\$ in \$\cH\N^{Id}\$
respectively are {\rm equivalent}
(physically the same observables) with
respect to the isomorphisms in Theorem
1:
$$A\N^{D,Q,sym}|_{\cH\N^{Id}}=
\Big(I\N^{D\rightarrow Id}\Big)A\N^D
\Big(I\N^{Id\rightarrow D}\Big).
\eqno{(10a)}$$ (One could write
pedantically
\$\Big(A\N^D|_{\cH\N^D}\Big)\$ instead
of simply \$A\N^D\$ in (10a).)

{\bf B)} Conversely, let \$B\N^{Id}\$
be any (completely) symmetric Hermitian
operator (identical-particle
observable) in \$\cH\N\$ that {\it
commutes} (is compatible with)
\$Q\N^{sym}\$ (cf (3b)). Then the
Hermitian operator (in \$\cH\N\$)
$$B\N^D\equiv\Big[(N!)\Big/\Big(\prod_{j=1}
^J(N_j!)\Big)\Big]Q\N
B\N^{Id}S\N^{s,a}Q\N \eqno{(10b)}$$
commutes with every distinct-cluster
permutation \$\forall p\in\cG_D:\enskip
[P\N ,B\N^D]=0,\$ and with \$Q\N .\$
The reducee of \$B\N^D\$ in \$\cH\N^D\$
is {\rm equivalent} with (physically
the same observable as) the reducee of
\$B\N^{Id}\$ in \$\cH\N^{Id}:\$
$$B\N^D|_{\cH\N^D}=\Big(I\N^{Id
\rightarrow D}\Big)
B\N^{Id}\Big(I\N^{D\rightarrow
Id}\Big).
\eqno{(10c)}$$}\\

Proof of Theorem 2 is given in Appendix
C.

The following result is an immediate
consequence of Theorem 1 and the two
parts of Theorem 2.\\

{\bf\noindent Corollary 1.} {\it If one
considers an {\it a priori} given
operator \$B\N^{Id}\$ as specified in
Theorem 2.B, and if one utilizes (10b)
to derive \$B\N^D\$ from it, and then
one takes the symmetrized form
$$B\N^{D,sym}\equiv \Big(\prod_{j=1}^J
(N_j!)\Big)^{-1}\sum_{p\in\cS_n}P\N
B\N^DP\N^{-1}$$ of the latter, then,
though, in general, \$B\N^{Id}\$ and
\$B\N^{D,sym}\$ are distinct operators
in \$\cH\N,\$ they have {\it one and
the same reducee} in \$\cH\N^{Id}.\$In
this sense, one can consider the latter
operator as rewriting the former
operator in suitable form (having
\$\cH\N^{Id}\$ in mind).}\\

{\bf\noindent Corollary 2.} {\it Let
\$\rho\N^{Id}\$ and \$ \rho\N^D\equiv
I\N^{Id\rightarrow
D}\rho\N^{Id}I\N^{D\rightarrow Id},\$
its equivalent distinct-cluster state
be given. Then
$$\Big<B\N^{Id}\Big>_{\rho\N^{Id}}\equiv
\tr\Big(\rho\N^{Id}B\N^{Id}\Big) =\tr
\Big(\rho\N^DB\N^D\Big) \eqno{(11)}$$
(cf Theorem 2.B), i. e., {\it the
expectation values are equal}.}\\

{\bf\noindent Proof} follows
immediately from (10c) and the fact
that the maps \$I\N^{D\rightarrow Id}\$
and \$I\N ^{Id\rightarrow D}\$ are
mutually inverse unitary isomorphisms
(cf Theorem 1).
\hfill $\Box$\\

One can speak of {\it
identical-particle representation}
(description in \$\cH\N^{Id}\$) and of
{\it distinct-cluster representation}
(treatment in \$\cH\N^D\$) in the sense
of Theorems 1, 2 and Corollary 2.\\

{\bf\noindent Corollary 3.} {\it Let
\$U\N^{Id}\$ be the {\it unitary
evolution operator} in \$\cH\N\$ for
some time interval for the
\$N$-identical-particle system under
consideration such that
\$[U\N^{Id},Q\N^{sym}]=0\$ (cf (3b),
\$\forall p\in\cS_N:\enskip
[U\N^{Id},P\N]=0,\$ and hence
\$[U\N^{Id},S\N^{s,a}]=0\$). Let,
further, \$\rho\N^{Id,i}\$ and
\$\rho\N^{Id,f}\$ be the initial and
the final \$N$-identical-particle
density operators (physically, states)
such that \$Q\N^{sym}\rho\N^{Id,k}=
\rho\N^{Id,k},\enskip k=i,f.\$ Then the
evolution
\$\rho\N^{Id,f}=U\N^{Id}\rho\N^{Id,i}
(U\N^{Id})^{-1}\$ can be transferred
from \$\cH\N^{Id}\$ to \$\cH\N^D\$ (cf
(5) and (4a-c)), i. e., the
identical-particle description can be
replaced by the distinct-cluster one,
to obtain
\$\rho\N^{D,f}=U\N^D\rho\N^{D,i}
(U\N^D)^{-1}.\$ (Naturally, this is due
to the unitary nature of the
isomorphisms \$I\N^{Id\rightarrow D}\$
and \$I\N^{D\rightarrow Id}\$ - cf (6)
and
(7)).}\\

Finally, as the last layer of quantum
mechanical description, we discuss
measurement. We confine ourselves to
{\it ideal measurement}, the one to
which most textbooks of \QM confine
themselves, and where the change of
state is given by the L\"{u}ders
formula.$^8$

Let \$A\N^{Id}\$ be an
\$N$-identical-particle Hermitian
operator (observable) in \$\cH\N\$ for
which \$[A\N^{Id},Q\N^{sym}]=0\$ (cf
(3b)) is valid. Let, further,
\$A\N^{Id}=\sum_i a_iE\N^i,\enskip
i\not= i'\enskip \Rightarrow\enskip
a_i\not= a_{i'}\$ be its (unique)
spectral form. As it is well known, the
commutation of \$A\N^{Id}\$ both with
\$P\N,\enskip \forall p\in\cS_N\$ and
with \$Q\N^{sym}\$ is necessarily valid
also for each eigen-projector \$E\N^i$.

{\it Nonselective} ideal measurement
converts any \$N$-identical-particle
state \$\rho\N\$ into the
(nonselective) L\"{u}ders state
$$\sum_i(E\N^i\rho\N E\N^i).\eqno{(12a)}$$

On the other hand, {\it selective}
ideal measurement, in which, e. g., the
fixed result \$a_i\$ that is
detectable, i. e., such that
\$\tr(E\N^i\rho\N)>0,\$ is selected,
converts \$\rho\N\$ into the selective
L\"{u}ders state
$$E\N^i\rho\N E\N^i\Big/
\Big(\tr(E\N^i\rho\N)
\Big).\eqno{(13a)}$$

After this elementary introduction, we
can reduce the given state changes to
the subspace \$\cH\N^{Id}\$ (cf (5)).
To this purpose, we make the assumption
that \$Q\N^{sym}A\N^{Id}\not= 0,\$ i.
e., that the operator \$A\N^{Id}\$ has
a nonzero reducee in \$\cH\N^{Id}\$
(physically, a relevant component) .
Further, we take a state possessing (cf
(D.2)) the distinguishing property:
\$Q\N^{sym}\rho\N^{Id}= \rho\N^{Id}.\$
Then the {\it nonselective L\"{u}ders
change of state} gives$^8$
$$\sum_i'\Big(E\N^i|_{\cH\N^{Id}}\Big)
\rho\N^{Id}\Big(E\N^i|_{\cH\N^{Id}}
\Big),\eqno{(12b)}$$ where the prim on
the sum denotes that the zero terms are
omitted. The {\it selective L\"{u}ders
change of state} reads$^8$ $$\forall
i,\enskip\tr(E\N^i \rho\N^{Id})>0:\quad
\rho\N^{Id}\quad\rightarrow$$ $$\Big(
E\N^i|_{\cH\N^{Id}}\Big)\rho\N^{Id}
\Big(E\N^i|_{\cH\N^{Id}}\Big)
\Big/\Big[\tr\Big((E\N^i|_{\cH\N^{Id}})
\rho\N^{Id}\Big)\Big].
\eqno{(13b)}$$\\

The isomorphism \$I\N^{Id\rightarrow
D},\$ mapping \$\cH\N^{Id}\$ onto
\$\cH\N^D,\$ further converts (12b) and
(13b) into
$$\sum_i'E\N^{i,D}\rho\N^D
E\N^{i,D},\eqno{(12c)}$$ and $$\forall
i,\enskip\tr(E\N^{i,D}
\rho\N^D)>0:\quad
\rho\N^D\quad\rightarrow\quad
E\N^{i,D}\rho\N^DE\N^{i,D}\Big/\Big(\tr
(E\N^{i,D}\rho\N^D)\Big).\eqno{(13c)}$$
respectively. Here $$E\N^{i,D}\equiv
\Big(I\N^{Id\rightarrow
D}\Big)E\N^i|_{\cH\N^{Id}}\Big(I\N^
{D\rightarrow Id}\Big).\eqno{(14)}$$\\

One should note that the isomorphism
\$I\N^{Id\rightarrow D},\$ which
enables one to decouple the clusters
from each other, applies only to {\it a
restricted set of observables}. This
set is determined by the requirement of
compatibility with the distinguishing
property \$Q\N^{sym}.\$Overmore, only
the reducees in \$\cH\N^{Id}\$
(physically, the relevant components)
are taken into account for transfer
into \$\cH\N^D\$ (physically, for
conversion into the distinct-cluster
description).
\pagebreak

\section{Illustrations}

{\bf\noindent \large A. Valence
electrons}

\noindent It is well known in quantum
molecular physics (also called quantum
chemistry) that only the outermost
so-called valence electrons, on which
the attractive action of the nucleus is
relatively weakest, partake in forming
the bonds between the atoms to make
molecules. Hence, to treat the bonds it
is practical to consider the core
electrons and the valence ones as
distinct particles. The distinguishing
properties are defined in terms of the
relevant shell-model single-particle
states.\\

\noindent {\large \bf B.
Non-overlapping spatial domains}

\noindent Let \$\cD_e\$ be a spatial
domain comprising a laboratory on
earth, and \$\cD_m\$ an analogous
domain on the moon. Let, further,
\$\cD_3\$ be any third spatial domain
disjoint with both preceding ones (and
these two are, of course, disjoint from
each other). The distinguishing
projectors are \$Q_i\equiv \int \int
\int_{\cD_i}\ket{\vec r}\bra{\vec r}
d^3\vec r,\enskip i=e,m,3.\$ They are
orthogonal due to the disjointness of
the domains.

Since all experiments are done in the
laboratories (on earth and on moon),
the relevant observables satisfy the
required restrictions of compatibility
with and preservation of the
corresponding distinguishing property.
Hence, the easier thing, and the thing
that is done, is to work in the
decoupled space \$\cH\N^D\$ and not in
the coupled space \$\cH\N^{Id}.\$More
on this in the critical discussion in
subsection 6.D.\\

\noindent {\large \bf C. Nucleons}

\noindent The single-nucleon state
space has three tensor-factor spaces:
the orbital (or spatial) one, the spin
one, and the isospin one. The
single-nucleon {\it distinguishing
projectors} are the eigen-projectors of
\$t_z,\$ the z-projection of isospin,
which is completely analogous to the
spin-$1/2\$ case. Protons correspond to
the eigenvalue \$t_z=+1/2\$ and neutron
to \$t_z=-1/2\$ respectively. (The
projectors are multiplied tensorically
by the identity operators in the
orbital and in the spin factor space).

When weak interaction does not play a
role, i. e., when no
\$\beta$-radioactivity (converting
protons into neutrons and {\it vice
versa}) is taking place, then the
distinguishing property \$Q\N^{sym}\$
is possessed by the state
\$\rho\N^{Id}\$ of the nucleus. Namely,
this property physically simply says
that there are \$N_p\$ protons and
\$N_n\$ neutrons in the \$N$-nucleonic
nuclear state \$\rho\N^{Id}\$
(\$N=N_p+N_n;\$ in nuclear physics the
notation is \$N_p=Z,\$ \$N_n=N,\$ and
\$N=A\$). Hence, one can transfer the
quantum-mechanical description from the
first-principle completely
antisymmetric space \$S\N^{s,a}\cH\N\$
(in which the so-called extended Pauli
principle is valid) to the effective
distinct-cluster space \$\cH\N^D.\$We
have two clusters here, that of protons
and that of neutrons. (Some formulae
are given in the preceding article$^3$
for the two-nucleon case, where the
deuteron is discussed in some detail.)
More about nucleons in subsection 6.C.\\

\section{Concluding remarks}

\noindent In this final section the
essential features of the expounded
theory are summed up, and some
important special
cases are critically discussed.\\

\noindent {\bf \large A. The effective
distinct-cluster subspace and its role}

\noindent The theory is based on
\$J\enskip \Big(2\leq J\leq N\Big)\$
{\it orthogonal single-particle
projectors} \$\{Q_1^j: j=1,2,\dots
,J\},\$ called "distinguishing
projectors". They determine the {\it
distinct-cluster subspace}
$$\cH_N^D\equiv\prod_{j=1}^{\otimes
J}\Big[S\M^{s,a}
\Big(\prod_{n=(M_j+1)}^{\otimes
(M_j+N_j)} (Q_n^j\cH_n)\Big)\Big]$$ (cf
 (4b)) of the (formal) distinct-particle
(most encompassing) space
\$\cH\N\equiv\prod _{n=1}^{\otimes
N}\cH_n.\$The subspace \$\cH_N^D\$ is
{\it isomorphic} (cf Theorem 1) to the
{\it corresponding
\$N$-identical-particle subspace}
\$\cH\N^{Id}\equiv Q\N^{sym}
\Big(S\N^{s,a}\cH\N\Big)\$ of the
first-principle identical-particle
state space \$S\N^{s,a}\cH\N\$ (cf
(5)). Here \$Q\N^{sym}\$ is the
symmetrized tensor product of the
distinguishing projectors (cf (3b) and
(3a)), called {\it the distiguishing or
the restricting property}.

The distinct-cluster subspace
\$\cH\N^D\$ is {\it relevant} only for
\$N$-identical-particle states
\$\rho\N^{Id}\$ that {\it possess} the
distinguishing property \$Q\N^{sym},\$
i. e., that satisfy
\$Q\N^{sym}\rho\N^{Id}=\rho\N^{Id}.\$
This key relation between subspace
\$\cH\N^{Id}\$ and state
\$\rho\N^{Id}\$ can, naturally, be
inverted: If an \$N$-identical-particle
state \$\rho\N\$ is given, and one can
find \$J\$ distinguishing
(single-particle) projectors making up
an (\$N$-identical-particle)
distinguishing property \$Q\N^{sym}\$
possessed by the state, then the
mentioned isomorphism (Theorem 1)
becomes relevant. ("Possession" of a
given property is explained in Appendix
D.)

In case of possession of the
distinguishing property by the state
\$\rho\N^{Id},\$ the mentioned
isomorphism
\$\cH\N^{Id}\enskip\rightarrow\enskip
\cH\N^D\$ enables one to make a
transition from \$\rho\N^{Id}\$  to a
corresponding effective
\$J$-distinct-cluster state, which is
denoted by \$\rho\N^D.\$Theorem 2
endows this transition with the
physical meaning of genuine (though
only effective) distinct-particle
clusters at the price of a serious
restriction: it is valid only for some
\$N$-identical-particle Hermitian
operators \$A\N\$ in \$\cH\N,\$ those
that {\it possess} the distinguishing
property \$Q\N^{sym}\$ (cf Appendix D):
they are Hermitian operators that {\it
commute} with the distinguishing
projector \$Q\N^{sym}\$ (physically,
that are compatible with the
distinguishing property), and of which
subsequently the reducee
\$A\N|_{\cH\N^{Id}}\$ in \$\cH\N
^{Id}\$ (physically, the relevant
component) is taken.

It is important to emphasize that the
effective distinct-cluster description
(in the effective state space
\$\cH\N^D,\$ cf (4a-c)) is {\it not an
approximation} (as effective particles
often are); for the observables that
possess the distinguishing property the
description is {\it exact}, and for
those that do not possess it, {\it it
does not make sense}.\\

\noindent
{\large \bf B) Converting
extrinsic properties into intrinsic
ones}\\

\noindent
As it was it was stated in
Definition 1, the notion of identical
particles rests on the idea of equal
intrinsic properties of the particles.
One can view the theory expounded as
the general framework how to {\it
convert some extrinsic properties},
represented by nontrivial projectors in
the single-particle state space, {\it
into intrinsic ones}. The converted
extrinsic properties are the
distinguishing projectors
\$\{Q_1^j:j=1,2,\dots ,J\}.\$In the
effective distinct-cluster space
\$\cH\N^D\$ these properties become
intrinsic (cf (4b)).\\

{\large \bf\noindent C) The reverse
algorithm: converting intrinsic
properties into extrinsic
ones}\\

\noindent Sometimes the {\it reverse
conversion} of intrinsic properties
into extrinsic ones takes place. For
this algorithm the same conceptual
framework can be used as for the direct
conversion (see the preceding section).
The theory presented in this article
covers also this case.

The best example is that of protons and
neutrons (cf subsection 5.C and the
preceding article$^3$).

If being a proton or a neutron is
considered as an intrinsic property of
the particle, then the state
\$\rho\N^D\$ of the nucleus is a
density operator in the
two-distinct-cluster space \$\cH\N^D
\equiv\Big(S_{1,\dots ,N_p}^a
\prod_{n=1}^{\otimes N_p}\cH_n^p\Big)
\otimes \Big(S_{1,\dots ,N_n}^a
\prod_{n=1}^{\otimes
N_n}\cH_n^n\Big),\$ where \$\cH_n^p\$
and \$\cH_n^n\$ are the space of the
\$n$-th single proton and the \$n$-th
single neutron. (The suffix \$n,\$
which denotes the neutron, should not
be confused with the index \$n\$.)

When weak interaction (or
\$\beta$-radioactivity) is taken into
account, the single-particle spaces
\$\cH_n^p\$ and \$\cH_n^n\$ have to be
replaced by a doubly dimensional {\it
nucleonic space} of the \$n$-th
particle \$\cH_n\equiv Q_n^p\cH_n\oplus
Q_n^n\cH_n,\$ where \$Q_n^p\$ and
\$Q_n^n\$ are the proton and the
neutron projectors respectively (cf
subsection 5.C), and \$\oplus\$ denotes
the orthogonal sum of subspaces. The
first-principle \$N$-identical-nucleon
space is then \$S\N^{s,a}\cH\N.\$
\$Q_n^p\$ and \$Q_n^n\$ are {\it the
distinguishing projectors}, and the
symmetrized
 \$N$-identical-nucleon projector
 \$Q\N\equiv\Big(
\prod_{n=1}^{\otimes N_p}Q_n^p\Big)
\otimes\Big(\prod_{n=(N_p+1)}^{\otimes
(N_p+N_n)}Q_n^n\Big)\$ is {\it the
distinguishing property} (cf (3a) and
(3b)). The corresponding
two-distinct-cluster space is now
rewritten as $$\cH\N^D
\equiv\Big(S_{1,\dots ,N_p}^a
\prod_{n=1}^{\otimes
N_p}Q_n^p\cH_n\Big)\otimes
\Big(S_{(N_p+1),\dots ,(N_p+N_n)}^a
\prod_{n=(N_p+1)}^{\otimes (N_p+N_n)}
Q_n^n\cH_n\Big),\eqno{(15)}$$ where
\$Q_n^p\cH_n=\cH_n^p\$ and \$Q_n^n\cH_n
=\cH_n^n\$ are the \$n$-th
single-proton and single-neutron spaces
respectively (cf subsection 5.C).

The reverse process at issue consists
in transferring the quantum-mechanical
description from \$\cH\N^D\$ to the
subspace \$\cH\N^{Id}\$ of the
first-principle space \$S\N^a\cH\N.\$
Inclusion of \$\beta$-radioactivity
requires the use of the latter space
because that of the
former does not suffice.\\

Perhaps additional light is shed on the
{\it reverse} application if the
expounded theory by discussing a {\it
fictitious case}. Suppose we want to
treat the proton (p) and the electron
(e) as two states of a single particle
(like the proton and the neutron). Can
we do this? The answer is affirmative,
and the way to do it is to use the
theory of this article in the, above
explained, reverse direction.

The new first-particle space would be
\$\cH_1\equiv Q_p\cH_1\oplus
Q_e\cH_1,\$ where \$Q_p\$ and \$Q_e\$
project \$\cH_1\$ onto the proton and
the electron subspace respectively. The
rest is analogous as in case of the
nucleon above with the important {\it
difference} that there is no
counterpart of the proton-state- or
neutron-state-property non-conserving
weak interaction. This means that every
\$N$-particle state \$\rho\N^{Id}\$
possesses the distinguishing property,
and can never lose it. Hence, the
corresponding distinct-cluster space
\$\cH\N^D\$ will always do for
description, and we are back to
permanently distinct
particles.\\

{\large \bf\noindent D) A critical
view of local quantum mechanics}\\

\noindent It was stated in subsection
6.A that the expounded theory is not an
approximate one; in some states and
some processes it is valid exactly, and
in others not at all. In some cases
approximation is nevertheless present
(in a different sense). Local \QM is
one of them.

For the description of, e. g.,
electrons in an earth laboratory (cf
subsection 5.B), the unsymmetrized
distinguishing property (cf (3a)) is
$$Q\N\equiv\Big(\prod_{n=1}^{\otimes
N_e}Q_n^e\Big)\otimes\Big(\prod_{n=
(N_e+1)}^{\otimes N}Q_n^{e\perp}
\Big)\eqno{(16)}$$ (cf (3a)), and
\$Q\N^{sym}\$ is its symmetrized form
(cf (3b)). In (16) \$N\$ and \$N_e\$
are the number of all electrons in the
universe and that of all electrons in
the earth domain \$\cD_e.\$The
projectors \$Q_n^e\$ and its
orthocomplementary \$Q_n^{e\perp}\$ map
the space \$\cH_n\$ onto the domain
\$\cD_e\$ and out of it respectively.

The corresponding distinct-cluster
space is
$$\cH\N^D\equiv\Big(S_{1,\dots ,N_e}
^a\prod_{n=1}^{\otimes
N_e}Q_n^e\cH_n\Big)\otimes\Big(S_{(N_e+1),
\dots ,N}^a\prod_{n=(N_e+1)}^ {\otimes
N}Q_n^{e\perp}\cH_n \Big).\eqno{(17)}$$

Since the two tensor factors in (17)
are {\it ipso facto} decoupled, one can
restrict the description to the first
factor \$\cH_{1,\dots ,N_e}^{Id}\equiv
S_{1,\dots ,N_e}
^a\Big(\prod_{n=1}^{\otimes
N_e}Q_n^e\cH_n\Big)\$ as far as
Hermitian operators (observables)
acting in this space are concerned.
Quantum-mechanical description in this
space we call {\it local \qm)}.

One wonders if there is anything wrong
with this. The answer is "yes". We give
two arguments against the exactness of
local \qm.\\

(i) We can imagine classically that
every electron is either on earth (in
\$\cD_e\$), or outside it. But
quantum-mechanically this is not so. In
any realistic state there are {\it
delocalized electrons}, which, put in a
simplified way, are in a state of
superposition of being on earth and
being outside it. Thus, the above
distinguishing property is not
possessed in an exact way.

This is where {\it approximation}
enters the scene. We approximate the
above realistic state by a state
\$\rho\N^{Id}\$ that possesses the
distinguishing property \$Q\N^{sym}\$
determined by (16) (cf (3a) and (3b)),
and we apply the presented theory to
it. All electrons that are involved in
laboratory experiments are certainly on
earth; the delocalized ones do not
participate. Hence the replacement of
the exact \$\rho\N\$ by the approximate
\$\rho\N ^{Id}\$ is believed to be a
good approximation.\\

(ii) When the orbital (or spatial)
tensor-factor space of a single
particle is determined by the basic set
of observables, which are the position,
the linear momentum, and their
functions, one obtains an {\it
irreducible} space, i. e., a space that
has no non-trivial subspace {\it
invariant simultaneously} for all the
basic observables (for position and
linear momentum; cf sections 5 and 6 in
chapter VIII of Messiah's book$^6$).
Hence, the above used subspace
\$Q_1^e\cH_1\$ (for the local, earth
quantum-mechanical description) is not
invariant either. It is, of course,
invariant for position, but linear
momentum has to be replaced by another
Hermitian operator approximating it.\\

{\large \bf\noindent  Appendix A.}\\

{\bf\noindent The mathematics required}\\

{\bf\noindent  A.I General properties
of groups and group representations
used in the article}.\\

{\bf \noindent General property A.I.1}
Multiplication of all elements \$g\$ of
a group \$\cG\$ from the left (or from
the right) maps \$\cG\$ in a one-to-one
way onto itself. (The claim is easily
proved.)\\

{\bf\noindent General property A.I.2}
Conjugation of all elements \$g\$ of a
group \$\cG\$ by any fixed element
\$g'\$ from the group maps \$\cG\$ in a
one-to-one way onto itself. Hence, \$
\{g'g(g')^{-1}:g\in \cG\}=\cG.\$ (An
immediate consequence of Property
A.I.1 .)\\

{\bf\noindent General property A.I.3}
If \$\cG'\$ is a proper finite subgroup
of a group \$\cG,\$ \$ g'\in \cG\$ is
an element outside \$\cG'\$ that leaves
\$\cG'\$ by conjugation invariant, then
the conjugation maps \$\cG'\$ in a
one-to-one way onto itself. (Immediate
consequence of the easily proved
one-to-one mapping and the finiteness
of \$\cG'.\$)\\

{\bf\noindent General property A.I.4}
Let \$\{x:x\in X\}\$ be a set, and let
\$\cG\$ be a group of transformations
acting in \$X.\$Further, let
\$\{f[x]:\forall f\}\$ be the set of
all single-valued mappings \$f\$ of
\$X\$ into some set \$Y\$ (\$f[x]\in
Y\$). Then \$\hat g f[x]\equiv
f[g^{-1}(x)]\$ induces a representation
of \$\cG\$ in the set of functions. One
should note that this means that the
map \$ \forall g\in \cG:\enskip
g\enskip \Rightarrow \enskip \hat g\$
is a {\it homomorphism}, i. e., it
preserves the product of two factors
due to
$$\hat g_1\hat g_2f[x]=\hat g_1f[g_2^{-1}(x)]=
f\Big[g_2^{-1}\Big(g_1^{-1}(x)\Big)\Big]=
f\Big[(g_1g_2)^{-1}(x)\Big]= \hat
{(g_1g_2)}f[x],$$ and it maps the unit
element \$e\$ into the unit element:
\$\hat ef[x]\equiv f[ex]=f[x]$.\\

{\bf\noindent General property A.I.5}
Let \$\cG\$ be a finite group of \$D\$
elements and \$\cG'\$ its subgroup of
\$d\$ elements. Subsets of the form
\$g\cG'\equiv\{gg':g'\in\cG'\}\$
($\cG'g$), \$g\in\cG,\$ are called left
(right) cosets of the subgroup. Let
\$\{g_k:k=1,2,\dots ,(D/d)\}\$ be
elements of \$\cG,\$ one from each left
coset, and symmetrically
\$\{g'_k:k=1,2,\dots ,(D/d)\}\$ from
the right cosets. Then
$$\sum_{k=1}^{D/d}g_k\cG'=\cG=
\sum_{k=1}^{D/d}\cG'g'_k\eqno{(A.1a,b)}$$
are the left-coset and the right-coset
(set-theoretical) decompositions of
\$\cG\$ into non-overlapping sets or
classes. (The symbol \$\sum\$ denotes
the union of disjoint sets. The left
and the right cosets coincide if and
only if the subgroup is an invariant
one.) Each coset contains \$d\$
elements, and one can take
\$g_{k=1}=g'_{k=1}\equiv e.\$ The first
cosets equal \$\cG'.\$(Property A.I.5
is a direct consequence of A.I.1 .)\\

{\bf\noindent General property A.I.6}
Taking the inverse element is an
anti-isomorphism (isomorphism with
transposing the factors) of any group
\$\cG\$ onto
itself. (Proof is straightforward.)\\

{\bf\noindent A.II Properties of the
group \$\cS_N\$ of all permutations
\$p\$ of \$N\$ objects used in the
article.}\\

{\bf\noindent Property A.II.1} Each
permutation \$p\in \cS_N\$ can be
factorized into transpositions, in
general, non-uniquely, but the number
of factors is either even in all
factorizations or odd. This property of
a permutation \$p\$ is unique, it is
written as \$(-)^p,\$ and it is called
"parity" of the permutation. It is by
definition \$+1\$ if the number of
factors is even, and it is $-1\$ in
case the number is odd. (See basic
ideas about groups in
Hamermesh$^9$ or in Messiah.$^6$)\\

{\bf\noindent Property A.II.2} Parity
\$(-)^p\$ of the permutation is a
homomorphism of \$\cS_N\$ into the
multiplicative group \$\{+1,-1\}:\$ \$
\forall p,p'\in \cS_N:\enskip
(-)^{pp'}=(-)^p (-)^{p'}.\$The unit
element \$e\$ has parity \$+1.\$\$
\forall p\in\cS_N:\enskip (-)^p=
(-)^{p^{-1}}\$ (as follows from \$
pp^{-1}=e\$).\\

{\bf\noindent Property A.II.3} The
group \$\cS_N\$ has only two
one-dimensional representations: the
so-called identical (or symmetric) one
(\$\forall p\enskip \Rightarrow 1\$)
and the antisymmetric one (\$\forall
p\enskip \Rightarrow (-)^p\$)$^6$  (p.
1117 there).\\

{\bf\noindent Property A.II.4} The
group \$\cS_N\$ has one and only one
invariant subgroup \$\cG':\$ the group
of even permutations
\$\cG'\equiv\{p:p\in \cS_N, (-)^p=1\}\$
(see Messiah's textbook$^6$,
p. 1111 there).\\

{\bf\noindent Property A.II.5} Let us
define for all \$p\in \cS_N\$ \$
sign(p)\equiv +1\$ if one deals with
bosons, and \$ sign(p)\equiv (-)^p\$ in
case of fermions. Further, let \$
\cG'\subseteq \cS_N\$ be any (proper or
improper) subgroup of \$\cS_N.\$ Let
\$\cG'\$ have \$d\$ elements. Finally,
we define \$ S\N^{\cG',s,a}\equiv
d^{-1}\sum_{p\in \cG}sign(p)P\N \$ (cf
A.II.6). One has
$$\forall p\in \cG':\quad
\Big(sign(p)P\N \Big)S\N^{\cG',s,a}
=S\N^{\cG',s,a} =S\N^{\cG',s,a}
\Big(sign(p)P\N \Big).
\eqno{(A.2a,b)}$$ (It is a direct
consequence of A.I.1 and A.II.2.)\\

{\bf\noindent Property A.II.6}  The
operator $S\N^{\cG',s,a}$ (cf A.II.5)
is a {\it projector}.

(This is so because
$$ (S\N^{\cG',s,a})^{\dagger}=
d^{-1}\sum_{p\in
\cG}sign(p)P\N^{\dagger}=
d^{-1}\sum_{p\in \cG}sign(p)P\N^{-1}=$$
$$ d^{-1}\sum_{p^{-1}\in
\cG}sign(p^{-1})P\N^{-1}=
S\N^{\cG',s,a}$$ - cf I.6 and II.2.
Further, $$(S\N^{\cG',s,a})^2=
d^{-1}\sum_{p\in \cG}sign(p)P\N
S\N^{\cG',s,a}=d^{-1}dS\N^{\cG',s,a}=
S\N^{\cG',s,a}$$ - cf (A.2a).)\\

{\bf\noindent Property A.II.7} The
symmetry projector $S\N^{\cG',s,a}$ of
any subgroup \$\cG'\$ of \$\cS_N\$ (cf
A.II.5 and A.II.6)) satisfies
$$S\N^{\cG',s,a}S\N^{s,a}=S\N^{s,a}=
S\N^{s,a}S\N^{\cG',s,a},\eqno{(A.3a,b)}$$
where
$$S\N^{s,a}\equiv S\N^{\cS_N,s,a}=(N!)^{-1}\sum_{p\in \cS_N}
sign(p)P\N.\eqno{(A.4)}$$

(Namely, on account of (A.2.a), one can
write $$S\N ^{\cG',s,a}S\N ^{s,a}=$$
$$d^{-1}\sum_{p\in \cG'} sign(p)P\N S\N
^{s,a}=d^{-1}\sum_{p\in \cG'}S\N
^{s,a}=\Big(d^{-1}d\Big)S\N ^{s,a}=S\N
^{s,a},$$ where \$d\$ is the order of
(the number of elements in) \$\cG'.\$
The symmetrical argument utilizing
(A.2.b) leads to
(A.3.b).)\\

{\bf\noindent A.III Basic properties of
the representations of \$\cS_N\$ in
\$\cH\N\$ that are used in the article.}\\

{\bf\noindent Property A.III.1} An
uncorrelated vector \$
\prod_{n=1}^{\otimes N}\ket{\psi_n}_n\$
in \$\cH\N\$ consists of the choice of
the \$N\$ vectors \$\ket{\psi_n}\$ from
the single-particle space (the \$N\$
spaces \$\cH_n\$ can be considered as
one thanks to the identicalness of the
particles - see the Introduction), and
of tensor multiplication (the second
index enumerates the factors). The
choice can be understood as a map of
the set \$\{1,\dots ,N\}\$ into the
single-particle space. Hence we are
dealing with \$f[x]\$ (cf A.I.4), and
we can apply the procedure of induction
specified in A.I.4 to obtain a
representation. Thus
$$\forall p\in \cS_N:\quad P\N
\Big(\prod_{n=1}^{\otimes N}
\ket{\psi_n}_n\Big)
=\prod_{n=1}^{\otimes
N}\ket{\psi_{p^{-1}(n)}}_n.\eqno{(A.5)}$$
(A.5) is a definition of the operators
\$P\N,\$ it is easily seen that it
coincides with that given in Section
3.\\

{\bf\noindent Property A.III.2} Let
\$\prod_{n=1}^{\otimes N}O_n^n\$ be the
tensor product of any \$N\$
single-particle operators \$
\{O^n:n=1,\dots ,N\}.\$Then
$$\forall p\in \cS_N:\quad P\N
\Big(\prod_{n=1}^{\otimes
N}O_n^n\Big)P\N^{-1}= \Big(\prod_{n=1}
^{\otimes N}O_n^{p^{-1}(n)}\Big).
\eqno{(A.6)}$$ (This is a
straightforward consequence of Property
A.III.1.)\\

{\bf\noindent Property A.III.3} Let \$
\prod_{n=1}^{\otimes N}O_n^n\$ be the
tensor product of \$N\$ single-particle
operators (with some possibly equal) \$
\{O^n:n=1,\dots ,N\}.\$Then (cf (A.4)):
$$S\N^{s,a}(\prod_{n=1}^{\otimes N}
O_n^n)S\N^{s,a}=(N!)^{-1}S\N^{s,a}\Big(\sum_{p\in
\cS_N}P\N(\prod_{n=1}^{\otimes N}
O_n^n)P\N^{-1}\Big)S\N^{s,a}.
\eqno{(A.7)}$$

(This is so because, making use of II.5 and II.2, one can write
$$\forall p\in \cS_N:\quad lhs= S\N^{s,a}P\N(\prod_{n=1}^{\otimes
N} O_n^n)P\N^{-1}S\N^{s,a}.$$ Adding this up for all
\$N!\$ permutations, (A.7) ensues.)\\

{\bf\noindent Property A.III.4} We now
write down a set-theoretical
decomposition that is relevant for the
distinct-cluster space \$\cH\N^D\$ (cf
(4a-c)).

Let \$N=\sum_{j=1}^JN_j,\$ \$\forall
j:\enskip 1\leq N_j\leq N\$ be a
decomposition of the natural \$N\$ into
\$J\$ naturals. Further, let \$\forall
j,\enskip j\geq 2:\enskip M_j\equiv
\sum_{j'=1}^{(j-1)}N_{j'};\$ and
\$M_{j=1}\equiv 0.\$ Finally, let
$$\{1,2,\dots
,N\}=\sum_{j=1}^J\{(M_j+1),(M_j+2),\dots
,(M_j+N_j)\}\eqno{(A.8)}$$ be a {\it
set-theoretical decomposition into
classes} (non-overlapping subsets -
hence the union is replaced by
\$\sum\$) each containing \$N_j\$
successive naturals. (The natural
\$M_j\$ is the number of naturals that
precede the j-th class.)

Let, further, \$j(n)\$ denote the class
to which \$n\$ belongs. We assume that
\$N\$ first-particle operators \$
\{O_1^n:n=1,\dots ,N\}\$ are given, and
that those and only those that belong
to one and the same class in the
decomposition (A.8) are equal. Hence,
the index \$j\$ enumerates also the
distinct operators, and one can rewrite
the set of operators as \$\{O_1^{j(n)}:
n=1,\dots ,N\}$.

The \$N$-particle operator (in
\$\cH\N\$)
$$O\N^{sym}\equiv\Big
(\prod_{j=1}^J(N_j!)\Big)^{-1}
\sum_{p\in \cS_N}P\N
\Big(\prod_{n=1}^{\otimes
N}O_n^{j(n)}\Big)P\N^{-1}\eqno{(A.9)}$$
is the {\it symmetrized} form of \$
\prod_{n=1}^{\otimes N}O_n^{j(n)}.\$It
is {\it symmetric}, i. e. \$ \forall
p\in \cS_N:\enskip
[O\N^{sym},P\N]=0\enskip
\Leftrightarrow \enskip P\N
O\N^{sym}P\N^{-1}= O\N^{sym}.\$(The
latter relation for \$O\N^{sym}\$
follows from I.1.)

The operator \$O\N^{sym}\$ consists of
\$(N!)\Big/\prod_{j=1}^J(N_j!)\$ {\it
distinct tensor-product terms}. To
prove this claim, we define \$\cG_D\$
as the subgroup of \$\cS_N\$ containing
all permutations that leave each class
in decomposition (A.8) invariant:
$$\forall p\in \cG_D,\enskip
\forall n:\quad
\mbox{if}\quad n\in \{(M_j+1),\dots ,(M_j+N_j)\}\quad \Rightarrow$$
$$ p(n)\in \{(M_j+1),\dots ,(M_j+N_j)\}.$$
Taking into account (A.1.a)
(\$\cG\equiv \cS_N,\enskip \cG'\equiv
\cG_D\$), (A.6), and using the
notations \$D\equiv (N!),\enskip
d\equiv \prod_{j=1}^J(N_j!)\$, the
definition (A.9) leads to
$$O\N^{sym}=d^{-1}
\sum_{p\in \cG_D}\sum_{k=1}^{D/d}
\prod_{n=1}^{\otimes N}
O_n^{j[(p_kp)^{-1}(n)]}=$$
$$d^{-1}
\sum_{p\in \cG_D}\sum_{k=1}^{D/d}
\prod_{n=1}^{\otimes N} O_n^{j[p^{-1}
(p_k^{-1}(n)]}.$$ Further, since the
permutation \$p^{-1},\$ \$ \Big(p\in
\cG_D\Big),\$ does not change the \$j\$
value of \$p_k^{-1}(n),\$ one further
has $$O\N^{sym}=(d^{-1}d)\sum_{k=1}
^{D/d} \prod_{n=1}^{\otimes N}
O_n^{j\Big(p_k^{-1}(n)
\Big)}=\sum_{k=1} ^{D/d}\prod_{n=1}
^{\otimes N}
O_n^{j\Big(p_k^{-1}(n)\Big)}.$$
Finally, arguing {\it ab contrario}, we
take \$k\not= k'\$ and assume that the
corresponding terms are equal: \$
\forall n:\quad
O_n^{j[p_k^{-1}(n)]}=O_n^{j[p_{k'}^{-1}(n)]}.\$
Then, \$ \forall n:\enskip
j[p_k^{-1}(n)]=j[p_{k'}^{-1}(n)]. \$
Hence, \$p_{k'}p_k^{-1}\equiv p\in
\cG_D, \enskip \Leftrightarrow \enskip
p_{k'}=pp_k\$ in contradiction with
(A.1.a).)\\

{\bf\noindent Property A.III.5} Let the
distinct operators \$O^j\$ in the
preceding property be {\it orthogonal
single-particle projectors} \$
\{Q^j:j=1,\dots ,J\}.\$ Then \$ Q\N
^{sym}\$ (cf (A.9)) is a symmetric {\it
projector} in \$\cH\N,\$ which consists
of \$\Big((N!)\Big/\prod_{j=1}^J(N_j!)
\Big)\$ orthogonal projector terms.

(This is due to the fact that now
"distinct" means orthogonal on the
single-particle level, hence the
product of any two distinct
tensor-product projector terms in
\$Q\N^{sym}\$ multiplies for some value
of \$n\$ orthogonal single-particle
projectors
giving zero.)\\

{\bf\noindent Property A.III.6} Let \$
O\N^D\$ be an operator in \$\cH\N\$
that commutes with all distinct-cluster
pemutations \$\forall p\in\cG_D:\enskip
[O\N^D,P\N]=0\$ (cf property A.III.4).
Then:
$$S\N^{s,a}O\N^DS\N^{s,a}=\Big[
\Big(\prod_{j=1}^J(N_j!)\Big)\Big/(N!)
\Big] S\N^{s,a}O\N ^{D,sym}
S\N^{s,a},\eqno{(A.10)}$$ where
$$O\N^{D,sym} \equiv
\Big(\prod_{j=1}^J(N_j!)\Big)^{-1}
\sum_{p\in
\cS_N}P\N\Big(O\N^D\Big)P\N^{-1}
\eqno{(A.11)}$$ is the symmetrized form
of \$O\N^D.\$

(This is so because, making use of
A.II.5 and A.II.2, one can write
$$\forall p\in \cS_N:\quad lhs(A.10)=
S\N^{s,a}P\N(O\N^D)P\N^{-1}S\N^{s,a}.$$
Adding this up for all \$(N!)\$
permutations, (A.10) is obtained.) The
operator \$O\N^{D,sym}\$ is symmetric.
(This is so because \$ \forall p'\in
\cS_N:\enskip P'\N
O\N^{D,sym}(P'\N)^{-1}=O\N^{D,sym}\$ as
follows from A.I.1.)\\

{\large \bf\noindent  Appendix B}\\

\noindent Now a {\bf proof of Theorem
1} is presented.

To begin with, we omit the restriction
sign \$|_{\dots}\$ wherever the
restriction is anyway fulfilled, and we
prove that \$ I\N^{Id\rightarrow D}
\cH\N^{Id}\subset \cH\N^D.\$One has (cf
(3a) and (3b)):
$$Q_{1\dots N}Q_{1\dots N}^{sym}=Q_{1\dots
N}=Q_{1\dots N}^{sym}Q_{1\dots
N},\eqno{(B.1a,b)}$$ due to
orthogonality of \$Q\N\$ to all the \$
\Big[(N!)\Big/\Big(\prod_{j=1}^J(N_j!)
\Big)\Big]\$ distinct projector terms
in \$Q_{1\dots N}^{sym}\$ except to the
term \$Q\N\$ itself (cf A.III.5). Also,
one has
$$\Big(\prod_{j=1}^{\otimes J}S\M
^{s,a}\Big)S\N^{s,a}=S\N^{s,a}
\eqno{(B.1c)}$$ (cf (A.3.a)).

Hence, in view of (6), (5), (B.1a), and
(B.1c), \$ I\N^{Id\rightarrow
D}\cH\N^{Id}= Q_{1\dots N}
\cH^{Id}_{1\dots N}= \Big[Q_{1\dots
N}\Big(\prod_{j=1}^{\otimes J}S\M
^{s,a}\Big)\Big]S^{s,a}_{1\dots
N}\cH_{1\dots N} .\$ In view of (4c),
this is a subspace of the space
\$\cH^D_{1\dots N}\$ because \$
\Big(S^{s,a}_{1\dots N}\cH_{1\dots N}
\Big)\$ is a subspace of \$\cH_{1\dots
N}$.\\

Next, we show the reverse claim that
the operator \$I\N^{D\rightarrow Id}\$
(given by (7)) takes the subspace \$
\cH^D_{1\dots N}\$ into \$
\cH^{Id}_{1\dots N}.\$ Utilizing (4c)
and (B.1b), one has
$$I\N^{D\rightarrow Id}\cH\N^D=S^{s,a}_{1\dots N}Q_{1\dots N}
\Big(\prod_{j=1}^{\otimes
J}S_{(M_j+1)\dots
(M_j+N_j)}^{s,a}\Big)\cH_{1\dots N}=$$
$$ S^{s,a}_{1\dots N}Q_{1\dots N}^{sym}Q_{1\dots
N}\Big(\prod_{j=1}^{\otimes
J}S_{(M_j+1)\dots (M_j+N_j)}^{s,a}\Big)
\cH_{1\dots N}.$$ Since \$
\Big[Q_{1\dots
N}\Big(\prod_{j=1}^{\otimes
J}S_{(M_j+1)\dots
(M_j+N_j)}^{s,a}\Big)\cH_{1\dots
N}\Big]\$ is a subspace of \$
\cH_{1\dots N}\$,  we have ended up
with a subspace of
\$ \cH^{Id}_{1\dots N}\$  (cf (5)).\\

Next, we show that the maps
\$I\N^{Id\rightarrow D}\$  and
\$I\N^{D\rightarrow Id}\$ in
application to the subspaces
\$\cH\N^{Id}\$ and to \$\cH\N^D\$
respectively are each other's inverse.

Owing to the definitions (7) and (6),
and to the definition (A.4) of
\$S\N^{s,a},\$ one has the following
equality of maps:
$$I\N^{Id\rightarrow D}I\N^{D\rightarrow
Id}=\Big\{\Big((N!)\Big/\prod_{j=1}
^J(N_j!) \Big)^{1/2}Q_{1\dots N}
\Big((N!)\Big/\prod_{j=1}^J(N_j!)\Big)^{1/2}
S^{s,a}_{1\dots N}\Big\}|_{\cH\N^D}=$$
$$\Big\{\Big(\prod_{j=1}^J(N_j!)
\Big)^{-1}\sum_{p\in
\cS_N}sign(p)Q_{1\dots N} P_{1\dots
N}\Big\}|_{\cH\N^D}.$$ On account of
(4c), we can further write
$$lhs=\Big\{\Big(\prod_{j=1}^J(N_j!)
\Big)^{-1} \sum_{p\in
\cS_N}sign(p)Q_{1\dots N} P_{1\dots
N}Q_{1\dots N}\Big\}|_{\cH\N^D}=$$
$$\Big\{\Big(\prod_{j=1}^J(N_j!)
\Big)^{-1} \sum_{p\in \cS_N}
sign(p)Q_{1\dots N}\Big(P_{1\dots
N}Q_{1\dots N} P_{1\dots
N}^{-1}\Big)P_{1\dots
N}\Big\}|_{\cH\N^D}.$$ All
\$\Big(P_{1\dots N}Q_{1\dots
N}P_{1\dots N}^{-1}\Big)\$ multiply
with \$Q_{1\dots N}\$ into zero except
when \$ p\in \cG_D \$ (cf A.III.4). For
the permutation operators from this
subgroup one has, \$ P_{1\dots
N}=\prod_{j=1}^{\otimes
J}P_{(M_j+1)\dots (M_j+N_j)},\$
corresponding (in the sense of
homomorphism) to \$ p\in \cG_D:\enskip
p=\prod_{j=1}^Jp_j,\$ where \$p_j\$
permutes possibly nontrivially only in
the \$j$-th class, in the rest of the
classes it permutes trivially (cf
(A.8)). Utilizing this, the idempotency
of \$Q_{1\dots N},\$ and the fact that
\$ Q_{1\dots N}\Big(\prod_{j=1}
^{\otimes J}S_{(M_j+1)\dots
(M_j+N_j)}^{s,a}\Big)\$ acts on
\$\cH_{1\dots N}^D\$ as the identity
operator, we can, further, write
$$lhs=\Big\{\Big(\prod_{j=1}^J(N_j!)
\Big)^{-1}\times$$ $$ \sum_{p\in
\cG_D}\Big[ Q_{1\dots
N}\Big(sign(p)\prod_{j=1} ^{\otimes
J}P_{(M_j+1)\dots
(M_j+N_j)}\Big)Q_{1\dots
N}\Big(\prod_{j=1}^{\otimes
J}S_{(M_j+1)\dots (M_j+N_j)}^{s,a}\Big)
\Big]\Big\}|_{\cH\N^D}. \eqno{(B.2)}$$

Further, one has \$
sign(p)=\prod_{j=1}^J sign(p_j)\$ for
\$p\in \cG_D\$ (cf A.II.2), and, in
accordance with (A.2.a), for each of
the \$J\$ class factors
$$sign(p_j)P_{(M_j+1)\dots
(M_j+N_j)}\Big(\prod_{n=(M_j+1)}^{\otimes
(M_j+N_j)}Q^j_n\Big) S_{(M_j+1)\dots
(M_j+N_j)}^{s,a}=$$
$$\Big(\prod_{M_j+1}^{\otimes
(M_j+N_j)}Q^j_n\Big) S_{(M_j+1)\dots
(M_j+N_j)}^{s,a}$$ is valid. Further,
using (3a) and (A.2a) once more, one
obtains
$$\Big(\prod_{j=1}^{\otimes J}sign(p_j) P_{(M_j+1)\dots
(M_j+N_j)}\Big)Q_{1\dots
N}\Big(\prod_{j=1}^{\otimes
J}S_{(M_j+1) \dots
(M_j+N_j)}^{s,a}\Big)=$$ $$ Q_{1\dots
N}\Big(\prod_{j=1}^{\otimes
J}S_{(M_j+1)\dots
(M_j+N_j)}^{s,a}\Big). \eqno{(B.3)}$$

Substituting (B.3) in (B.2), and
recognizing that the sum \$ \sum_{p\in
\cG_D}\$ adds up precisely \$
\Big(\prod_{j=1}^J(N_j!)\Big)\$ equal
terms, we finally have
$$lhs=\Big\{Q_{1\dots
N}\Big(\prod_{j=1}^{\otimes
J}S_{(M_j+1)\dots
(M_j+N_j)}^{s,a}\Big)\Big\}|_{\cH\N^D}
=1|_{\cH\N^D}.$$

This establishes the claim that
\$I\N^{Id\rightarrow D}\$ is the
inverse of \$I\N^{D\rightarrow Id}$.\\

Analogously, in view of (7), (6), and
(A.4), we have the following equality
of maps:
$$I\N^{D\rightarrow Id}I\N^{Id\rightarrow
D}=\Big\{\Big((N!)\Big/
\prod_{j=1}^J(N_j!)
\Big)^{1/2}S_{1\dots N}^{s,a}
\Big((N!)\Big/
\prod_{j=1}^J(N_j!)\Big)^{1/2}
 Q_{1\dots N}\Big\}|_{\cH\N^{Id}}=$$
$$\Big\{\Big(\prod_{j=1}^J(N_j!)\Big)^{-1}
\sum_{p\in \cS_N}\Big(P_{1\dots N}
Q_{1\dots N}P_{1\dots
N}^{-1}\Big)sign(p) P_{1\dots
N}\Big\}|_{\cH\N^{Id}}.$$ In view of
(5), in \$\cH_{1\dots N}^{Id}\$ \$
S_{1\dots N}^{s,a}\$ acts as the
identity operator. Therefore, taking
into account (A.2.a) and (3b), one
further obtains
$$ lhs=\Big\{\Big(\prod_{j=1}^J(N_j!)
\Big)^{-1}\sum_{p\in
\cS_N}\Big(P_{1\dots N} Q_{1\dots
N}P_{1\dots N}^{-1}\Big)sign(p)
P_{1\dots
N}S\N^{s,a}\Big\}|_{\cH\N^{Id}}=$$
$$\Big\{Q\N^{sym}S\N^{s,a}\Big\}
|_{\cH\N^{Id}}=1|_{\cH\N^{Id}}.$$ Thus,
the claim that the two maps are the
inverse of each other is proved.

Since the maps are the inverse of each
other, it is easily seen hat they are
necessarily surjections and injections,
i. e., bijections as claimed.\\

Next, we prove that \$I\N^{D\rightarrow
Id}\$ preserves the scalar product,
which we write as \$\Big(\dots ,\dots
\Big).\$Let \$\Psi\N\$ and \$\Phi\N\$
be two arbitrary elements of
\$\cH_{1\dots N}^D.\$On account of (7)
and the Hermiticity and the idempotency
of \$\S\$ (cf (A.4) and A.II.6), one
has
$$\Big(I\N^{D\rightarrow Id}\Psi\N
,I\N^{D\rightarrow Id}\Phi\N\Big)=$$
$$\Big(\6^{1/2}\Psi\N ,\6^{1/2}S\N^{s,a}
\Phi\N\Big)=$$
$$\6\Big(\Psi\N ,S\N^{s,a}\Phi\N\Big).$$
Further, on account of (A.4) and the
fact that \$Q\N\$ acts as the identity
operator on \$\Psi\N\$ and \$\Phi\N,\$
one can apply it to both and write
$$lhs=\Big(\prod_{j=1}^J(N_j!)\Big)^{-1}
\sum_{p\in \cS_N} sign(p)\Big(\Psi\N
,Q\N (P\N Q\N P\N ^{-1})P\N
\Phi\N\Big).$$ One has \$ Q\N (P\N Q\N
P\N ^{-1})=0,\$ except if \$ p\in
\cG_D,\$ when it is equal to \$Q\N.\$
If \$ p\in \cG_D,\$ then \$p\$ permutes
only within the classes, hence the
corresponding permutation operator
\$P\N\$ commutes with \$Q\N\$ (cf
(3a)). Also \$ \Phi\N
=\Big(\prod_{j=1}^{\otimes
J}(S_{(M_j+1)\dots
(M_j+N_j)}^{s,a})\Big)\Phi\N\$ (cf
(4c)) is valid, and
 for \$p\in \cG_D,\$ \$ sign(p)P\N
 \Big(\prod_{j=1}^{\otimes
J}(S_{(M_j+1)\dots
(M_j+N_j)}^{s,a})\Big)=\prod_{j=1}^{\otimes
J} (S_{(M_j+1)\dots (M_j+N_j)}^{s,a})\$
(cf A.II.5). Therefore,
$$lhs= \Big(\prod_{j=1}^J(N_j!)\Big)^{-1}
\sum_{p\in \cG_D}\Big(\Psi\N ,\Phi\N
\Big)= \Big(\Psi\N ,\Phi\N\Big).$$

It is easy to see that also the inverse
of a scalar-product preserving
bijection must be scalar-product
preserving. This completes the
proof of Theorem 1.\hfill $\Box$\\

{\large \bf\noindent  Appendix C}\\

\noindent {\bf We prove now Theorem 2}.

{\bf\noindent A)} To prove that the
operator
$$A\N^{D,Q,sym}\equiv\Big(\prod_{j=1}
^J(N_j!)\Big)^{-1}\sum_{p\in\cS_N}P\N
A\N^DQ\N P\N^{-1}$$ (cf (9)) commutes
with \$Q\N^{sym},\$ we utilize the
idempotency of \$Q\N\$ and its
commutation with \$A\N^D,\$ and we
rewrite \$A\N^{D,Q,sym}\$ in the form
\$A\N^{D,Q,sym}=\Big(\prod_{j=1}
^J(N_j!)\Big)^{-1}\sum_{p\in\cS_N}P\N
Q\N A\N^DQ\N P\N^{-1}.\$ Then, on
account of the facts that \$Q\N^{sym}\$
is symmetric and that
\$Q\N^{sym}Q\N=Q\N Q\N^{sym}=Q\N\$ (cf
(B.1a,b)), the claimed commutation
becomes obvious.

Next, to prove (10a), we start with its
rhs and we utilize (7) and (6):
$$rhs(10a)\equiv\Big(I\N^{D\rightarrow Id}\Big)
A\N^D\Big(I\N^{Id\rightarrow D}\Big)=$$
$$\Big[(N!)\Big/\Big(\prod_{j=1}^J(N_j!)\Big)
\Big]\Big[S\N^{s,a}A\N^DQ\N
\Big]|_{\cH\N^{Id}}.$$ Since
$S\N^{s,a}$ acts as the identity
operator in \$\cH\N^{Id},\$ one further
has
$$rhs(10a)=\Big[(N!)\Big/
\Big(\prod_{j=1}^J(N_j!)\Big)
\Big]\Big[S\N^{s,a}A\N^DQ\N
S\N^{s,a}\Big]|_{\cH\N^{Id}}.$$

Finally, on account of (A.10),
$$rhs(10a)=\Big[S\N^{s,a}A\N^{D,Q,sym}S\N
^{s,a}\Big]
|_{\cH\N^{Id}}=A\N^{D,Q,sym}|_{\cH\N^{Id}}
\equiv lhs(10a).$$\\

{\bf B)} It is obvious that the
operator $$B\N^D\equiv\Big[(N!)\Big/
\Big(\prod_{j=1}^J(N_j!)\Big)\Big] Q\N
B\N^{Id}S\N^{s,a}Q\N$$ (cf (10b))
commutes both with every
distinct-cluster permutation and with
\$Q\N .\$

Proof of (10c) is straightforward:
Utilizing the definitions (6) and (7)
of the isomorphisms, one obtains
$$rhs(10c)=
\Big[(N!)\Big/\Big(\prod_{j=1}^J(N_j!)\Big)
\Big]Q\N B\N^{Id}\Big[S\N^{s,a}\Big]|_
{\cH\N^D}.$$ Since in \$\cH\N^D\$
\$Q\N\$ acts as the identity operator,
a glance at (10b) establishes the
claim.\hfill $\Box$\\

{\large \bf\noindent Appendix D}\\

\noindent The mathematical relation
$$A\rho =a\rho ,\eqno{(D.1)}$$ where
\$A\$ is a Hermitian operator, \$\rho\$
a density operator, and \$a\$ a real
number, implies that \$a\$ belongs to
the eigenvalues of \$A,\$ and that
\$\rho\$ is a state in which the system
has the eigenvalue \$a\$ of \$A.\$ This
can be seen by substituting a spectral
form \$ \rho = \sum_ir_i\ket{\psi
}_i\bra{\psi }_i\$ in (D.1), and by
multiplying scalarly the obtained
relation from the right by one of the
eigenvectors of \$\rho ,\$ e. g., by
\$\ket{\psi }_{\bar i},\$ that
corresponds to a positive eigenvalue
\$r_{\bar i}.\$ Then \$ A\ket{\psi
}_{\bar i}=a\ket{\psi }_{\bar i}\$
ensues. (It is easy to extend the
argument to any decomposition of
\$\rho\$ into pure states by expanding
these in the complete eigenbasis, and
by taking into account that expansion
coefficients along eigenvectors of
\$\rho\$ corresponding to zero are
zero.)

If the observable \$A\$ is a property
(projector) \$F,\$ then one has the
following special case of (D.1):
$$F\rho =\rho .\eqno{(D.2)}$$ It has the
physical meaning that the system in the
state \$\rho\$ {\it possesses the
property} \$F.\$Namely, the probability
of \$F\$ in \$\rho\$ is \$1,\$ i. e.,
in suitable measurement, the event
\$F\$ necessarily occurs (the property
\$F\$ is necessarily obtained) in the
state \$\rho.\$

It is immediately seen (by adjoining)
that (D.2) implies
$$[\rho ,F]=0 \eqno{(D.3)}$$ with the
physical meaning of {\it compatibility}
of state and property.

Relation (D.3) expresses a weaker
property of the state \$\rho\$ than
relation (D.2). If only the former is
valid, then \$F\rho F=F\rho\$ satisfies
also the latter relation.\\

Besides (D.1), there is another
mathematical generalization of (D.2)
relevant for the investigation in this
article:
$$FA=A,\eqno{(D.4)}$$ where \$F\$ is a
projector, and \$A\$ is a Hermitian
operator (an observable). Like above,
(D.4) implies the (weaker)
compatibility relation $$[F,A]=0.
\eqno{(D.5)}$$ Conversely, if a
Hermitian operator \$B\$ satisfies only
(D.5) {\it mutatis mutandis}, i. e., if
\$[F,B]=0,\$ then for the Hermitian
operator (observable) \$FB\$ the
stronger relation (D.4) is valid {\it
mutatis mutandis}:
$$F(FB)=(FB) \eqno{(D.6)}$$ (as obvious
due to the idempotency of \$F\$).

If \$A=\sum_ia_iP_i,\enskip i\not=
i'\enskip\Rightarrow\enskip a_i\not=
a_{i'}\$ is the (unique) spectral
decomposition of \$A,\$ then (D.4)
implies $$\forall i,\enskip a_i\not=
0:\quad FP_i=P_i.\eqno{(D.7a)}$$ (This
is seen by substituting the spectral
form of \$A\$ on both sides of (D.4),
and by multiplying subsequently the
relation by \$P_i\$ - with a fixed
value of \$i.\$Taking into account that
(D.5) is, as well known, equivalent to
\$\forall i:\enskip [F,P_i]=0,\$ the
relation \$a_iFP_i= a_iP_i\$ ensues.)

As well known, (D.7a) is symbolically
written as $$\forall i,\enskip a_i\not=
0:\quad P_i\leq F\eqno{(D.7b)}$$ with
the physical meaning that if the event
\$P_i\$ {\it occurs}, i. e., the result
\$a_i\not= 0\$ is obtained in a
measurement of the observable \$A,\$
then {\it necessarily} also the event
\$F\$ {\it occurs} (or the property
\$F\$ is valid).

For the null projector of a Hermitian
operator \$A\$ that satisfies (D.4), i.
e., for the eigen-projector of the
latter corresponding to the eigenvalue
zero, only the weaker condition of
compatibility with \$F\$ is valid.
Hence, as it is easy to see, instead of
(D.7b) one has
$$a_{i_0}=0\quad\Rightarrow\quad
P_{i_0}=FP_{i_0}+F^{\perp},
\eqno{(D.7c)}$$ where \$F^{\perp}\equiv
1-F.\$

In the sense of (D.7b) and (D.7c), one
can give (D.4) {\it the physical
interpretation} that {\it the
observable \$A\$ possesses the
property} \$F.\$

It is perhaps interesting to realize
that {\it geometrically} (D.4) means
that the (topologically closed) range
\$\bar\cR(A)\$ of \$A\$ is entirely
within that of \$F:\$
\$\bar\cR(A)\subseteq\cR(F).\$In more
detail, \$\forall i,a_i\not= 0:\enskip
\cR(P_i)\subseteq\cR(F),\$ and for
\$a_{i_0}=0,\$ \$\cR(FP_{i_0})\subseteq
\cR(F),\enskip\cR(F^{\perp}P_{i_0})=
\cR(F^{\perp})\$ (cf (D.7c)).\\

{\bf \large\noindent References}

\noindent 1. W. M. De Muynck, {\it Int.
J. Theor. Phys.} {\bf 14} 327 (1975).

\noindent 2. L. I. Schiff,  {\it
Quantum Mechanics}, Chap. 9.
(McGraw-Hill Inc., New

\setlength{\parindent}{3ex} York,
1955).

\noindent 3. F. Herbut,  {\it Am. J.
Phys} {\bf 69} 207 (2001).

\noindent 4. R. Mirman,  {\it Nuovo
Cimento} {\bf 18B} 110 (1973).

\noindent 5. J. M. Jauch, {\it
Foundations of Quantum Mechanics},
Chap. 15.

\setlength{\parindent}{3ex}
(Addison-Wesley, Reading,
Massachusetts, 1966).

\noindent 6. A. Messiah, {\it Quantum
Mechanics} (North-Holland, Amsterdam,
1961).

\noindent 7. C. Cohen-Tannoudji, B.
Diu, and F. Laloe,  {\it Quantum
Mechanics},

\setlength{\parindent}{3ex} (John Wiley
and Sons, New York, 1977).

\noindent 8. G. L\"{u}ders,  {\it Ann.
Phys.} (Leipzig) {\bf 8} 322 (1951).
(In textbooks: p. 333

\setlength{\parindent}{3ex} in
Messiah's book, and p. 221 (Fifth
Postulate) in that of Cohen-

\setlength{\parindent}{3ex} Tannoudji
et al.)

\noindent 9. M. Hamermesh,  {\it Group
Theory} (Addison-Wesley, London, 1964).

\end{document}